\documentclass[twocolumn,showpacs,preprintnumbers,amsmath,amssymb]{revtex4}

\usepackage{graphicx}
\usepackage{dcolumn}
\usepackage{bm}
\usepackage{psfrag}

\newcommand{\be}{\begin{equation}}  
\newcommand{\ee}{\end{equation}}  
\newcommand{\bear}{\begin{eqnarray}}  
\newcommand{\eear}{\end{eqnarray}}  
\newcommand{\ba}{\begin{array}}  
\newcommand{\ea}{\end{array}}

\newskip\humongous \humongous=0pt plus 1000pt minus 1000pt

\newif\ifdtup

\def\oldreffmt#1{\rlap{[#1]} \hbox to 2\parindent{}}

\def\figfmt#1{\rlap{Figure {#1}} \hbox to 1in{}}  
  
\def\ie{\hbox{\it i.e.}{}}      \def\etc{\hbox{\it etc.}{}}  
\def\eg{\hbox{\it e.g.}{}}

\def\Tr{\mathop{\rm Tr}}

\def\slash#1{#1\!\!\!/\!\,\,}  
\def\beq{\begin{equation}}  
\def\eeq{\end{equation}}  
\def\bea{\begin{eqnarray}}  
\def\eea{\end{eqnarray}}

\def\bq{\begin{quote}}  
\def\eq{\end{quote}}

\relax  

\newdimen\tdim  
\tdim=\unitlength  
\def\bar{\overline}

\begin{document}

\preprint{EFI Preprint 07-21}
\preprint{FERMILAB-Pub-07/401-T}
\title{
Anomaly Mediated Neutrino-Photon Interactions\\
at Finite Baryon Density}

\author{Jeffrey A. Harvey$^{(a)}$}

\author{Christopher T. Hill$^{(b)}$}

\author{Richard J. Hill$^{(b)}$}

\affiliation{\vspace{0.2in}
$^{(a)}$Enrico Fermi Institute and Department of Physics \\
The University of Chicago, Chicago, Illinois, 60637, USA\\ 
$^{(b)}$Fermi National Accelerator Laboratory\\
\it P.O. Box 500, Batavia, Illinois, 60510, USA
\vspace{0.1in}
}


\begin{abstract} 
We propose new physical processes based on the axial vector anomaly
and described by the Wess-Zumino-Witten term that couples the photon,
$Z$-boson, and the $\omega$-meson.  The interaction takes the form of
a pseudo-Chern-Simons term, $\sim
\epsilon_{\mu\nu\rho\sigma}\omega^\mu Z^\nu F^{\rho\sigma}$.  This
term induces neutrino-photon interactions at finite baryon density via
the coupling of the $Z$-boson to neutrinos.  These interactions may be
detectable in various laboratory and astrophysical arenas.  The new
interactions may account for the MiniBooNE excess.  They also produce
a competitive contribution to neutron star cooling at temperatures
$\gtrsim 10^9\,{\rm K}$.  These processes and related axion--photon
interactions at finite baryon density appear to be relevant in many
astrophysical regimes.
\end{abstract}

\pacs{11.15.-q,12.15.-y,12.38.Qk,12.39.Fe,13.15.+g,13.40.-f,14.70.Hp,14.80.Mz,95.85.Ry,97.60.Jd}

\maketitle

\section{Introduction}

The axial vector anomaly plays a fundamental role in the 
structure of the Standard Model and
describes many physical processes, including the classic decay 
$\pi^0\rightarrow 2\gamma$
\cite{BJ,Adler,Bardeen}.
One can summarize the traditional current algebra manipulations
used to treat anomalous
processes by an effective action.  This is a functional,
$\Gamma(U,A_L,A_R)$, of a chiral field of the pseudoscalar mesons,
$U=\exp(i\pi^a t^a/f_\pi)$, and background gauge fields $A_L$, $A_R$, coupled
to left- and right-handed chiral currents. It
generates (consistent)
anomalies under local gauge transformations
and is known as the Wess-Zumino-Witten (WZW) term \cite{Wess,Witten}.

The WZW term has been developed into a phenomenologically
useful form by Kaymakcalan, Rajeev and Schechter (KRS) \cite{KRS}
following Witten's pioneering work. It 
can be understood as arising from a Chern-Simons term  built
of Yang-Mills gauge fields in $D=5$, suitably compactified such that
$A_5$ zero modes emerge as the mesons \cite{cth}. It applies to any effective
theory of pseudo-Nambu-Goldstone bosons (pNGB's) coupled 
to gauge fields, \eg, as in the case
of Little Higgs theories \cite{Hill2}. 
The WZW term arises naturally in connection
with topological physics in extra dimensions, 
and occurs in both top-down \cite{Sakai} and bottom-up
\cite{Erlich, DaRold} approaches to holographic QCD in order to correctly match
the flavor anomalies of QCD. 

The WZW term for spontaneously broken $SU(3)_L\times SU(3)_R$ flavor symmetry
describes anomalous processes involving the pseudoscalar pNGB's
and fundamental gauge fields such as the photon.  
For example, we have $\pi^0, \eta \rightarrow 2\gamma$
with $A_L = A_R = A Q$ and $Q={diag}(2/3, -1/3, -1/3)$.
However, the WZW term also  
summarizes processes containing effective vector mesons
in pole approximation, such as $\Phi\rightarrow 3 \pi$,
$\Phi\rightarrow K\bar{K}$, where $A_L = A_R = \Phi\lambda^8/2$;
and  $\omega\rightarrow \rho\pi\rightarrow 3\pi$, 
$\omega\rightarrow \pi^0\gamma$, \etc., where $A_L = A_R = 
\omega B + \rho I_3$, with $I_3={diag}(1/2,-1/2,0)$ and 
$B = {diag}(1/3,1/3,1/3)$  \cite{KRS}.

To leading order in an expansion in $\pi^a$ the WZW term 
is seen to contain
``pseudo-Chern-Simons'' terms (pCS)~\footnote{We use the terminology
``pseudo-Chern-Simons term'' to distinguish from ``Chern-Simons terms''
which only occur in odd spacetime dimension.},
such as 
$\Tr(\epsilon_{\mu\nu\rho\sigma} A^\mu_{L}A^\nu_R \partial^\rho A^\sigma_L)$. 
Recently it has been proposed that 
terms involving $A_1\omega d\rho$ may be of phenomenological interest
\cite{Domokos},
which has stimulated the present work.

Presently we note that the Standard Model implies a pCS term in the 
Lagrangian involving the photon,
the Z-boson, and the isoscalar $\omega$ vector meson 
of the form:
\beq
\label{one}
\frac{N_c}{48\pi^2} {e g_\omega g_2 \over \cos\theta_W} 
\epsilon_{\mu\nu\rho\sigma}
\omega^\mu Z^\nu F^{\rho\sigma} \,. 
\eeq
The derivation of this result 
requires a careful accounting of anomalies.
In the absence of $\omega$, gauge invariance 
is maintained by the combination of the WZW term
and anomaly cancelling contributions from the lepton sector. 
In the presence of $\omega$, $SU(2)_L \times U(1)_Y$  gauge
invariance must be enforced 
by including appropriate counterterms,  
$\epsilon_{\mu\nu\rho\sigma}A_L^\mu A_R^\nu \partial^\rho A_L^\sigma$, \etc. 
This uniquely specifies 
the coefficient of eq.\eqref{one}, apart from coupling
constant normalizations.  
$Z_\mu$ in eq.\eqref{one} should be thought of as a gauge-invariant 
Stueckelberg field, \ie, we are in unitary gauge 
for the broken $SU(2)_L\times U(1)_Y$
generators.
For simplicity, we restrict attention
to two light flavors and
the $\omega$ is 
coupled to baryon number $B={diag}(1/3,1/3)$, inducing
the coupling
to the nucleon isodoublet $N =(p,n)$, 
as $g_\omega \omega^\mu \bar{N}\gamma_\mu N$~%
\footnote{
Upon introducing $\omega$ in this way,
the WZW term contains a term 
$g_\omega \omega^\mu J_\mu$, where $J_\mu$ is the properly
normalized Goldstone-Wilczek Skyrmion baryon current when $N_c=3$.}. 
We also note that the one-loop diagram responsible for the pCS term
is closely related to the electroweak baryon
number anomaly. The full details will be presented
elsewhere \cite{Harvhill2}.

\begin{figure}[t]  
\vspace{2.0cm}  
\includegraphics{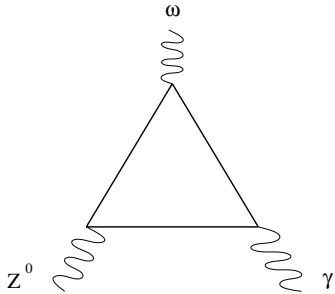}  
\vspace{2.25cm}  
\caption[]{
\addtolength{\baselineskip}{-.3\baselineskip}   
Triangle diagram at chiral constituent 
quark level treating $\gamma$, $Z$
and $\omega$ as background gauge fields.
This is contained in the Wess-Zumino-Witten term,
$\Gamma(U,A_L,A_R)$, which is an effective action expressing 
the full anomaly physics
of pseudoscalars, axial- and vector mesons 
and fundamental gauge fields. 
Integrating out $Z$ and $\omega$
leads to a neutrino-photon interaction at finite baryon density.}  
\label{triangle}  
\vspace{-0.25cm}
\end{figure}

Physically interesting effects arise when we integrate
out the $\omega$ and the $Z$, replacing them with the
baryon current and neutrino current respectively.
The new anomalous interaction is then:
\beq
S_{\rm int} = \sum_{f} \kappa \int d^4x\; \epsilon^{\mu\nu\rho\sigma} J_\mu \, 
\bar{\nu}^f_{L}\gamma_\nu \nu^{f}_L F_{\rho\sigma} \,,
\eeq
where $J_\mu =  \bar{N}\gamma_\mu N$ is the isosinglet baryon current 
and we sum over all left-handed neutrino flavors.
The normalization factor is:
\beq
 \kappa = \frac{N_c}{12 \pi^2}\frac{g_\omega^2 }{m_\omega^2}
 \frac{eG_F}{\sqrt{2}} \,, 
 \qquad \nu_L =\frac{1-\gamma^5}{2}\nu \,.
\eeq
One could include similar terms for the $\rho$ and $\Phi$ mesons,
but $\omega$ is expected to dominate.

Interesting effects will arise in processes such as:
\bea
\nu + N & \rightarrow & \nu + \gamma + N\; (or\; N')
\\
\gamma + N & \rightarrow & \nu + \bar{\nu} + N\; (or\; N') \,. 
\eea
Coherent enhancements to these basic processes can occur
in situations described by an ambient background baryon density. 

We presently turn to two such classes of phenomena: 
neutrino scattering on nucleons; and 
the cooling of a young neutron star. 
We briefly mention that an analogous
axion-photon-baryon current interaction arises
with similarly important implications.

\section{Neutrino--Nucleon Collisions}

We consider an incident neutrino
of any flavor with 
4-momentum $p=(E_\nu, \vec{p})$ colliding with a
stationary nucleon, producing an outgoing neutrino of momentum $k$, 
and a photon of momentum $q=(E_\gamma, \vec{q})$. 
We assume $E_\nu \lesssim M$, where
$M$ is the nucleon mass and recoil is treated
as negligible.
The action then becomes:
\beq
\label{result2}
S_{int}
\to \kappa \int d^4x\; 
\epsilon_{0ijk}\bar{\nu}_{L} \gamma^i \nu_L F^{jk}\; \delta^3(\vec{x}) \,. 
\eeq
The resulting total cross-section is: 
\bea\label{result}
\sigma & = & 
\frac{\alpha g^4_{\omega} G^2_F }{480 \pi^6 m_\omega^4}{E_\nu^6} 
 \nonumber \\
& = & 
2.6\times 10^{-41} {(E_\nu/{\rm GeV})^6} (g_\omega/10.0 )^4\; {\rm cm}^{2} \,. 
\eea
We remark that there is considerable uncertainty in the value of
$g_\omega$, generally extracted from nuclear potential models
\cite{Machleidt}, and the value we use here is ``conservative.''

To assess the possible experimental sensitivity to this effect 
we note that
MiniBooNE observes an excess
of $\sim 10^2$ events with electromagnetic energy
and with reconstructed mean neutrino
energies of order $\sim 400$ MeV \cite{Miniboone,Conrad}.
These events look like $400$ MeV $\nu_e+N\rightarrow e+N'$ charged current
events, but the electron could be faked by the hard photon in our process.

The MiniBooNE $\nu_\mu$ beam spectrum is fairly flat
from $300$ MeV to $1\,{\rm GeV}$, peaking at $\sim 700$ MeV.
We focus on $E_\nu \sim 700$ MeV $\nu_\mu$'s 
and assume that these 
produce photons with $E_\gamma \sim 400$ MeV via our process. 
In this energy range MiniBooNE accepts $\sim 2\times 10^5$
charged current quasi-elastic (CCQE) $\nu_\mu N\rightarrow \mu N^\prime$ 
events with cross-section,
$\sigma_{QE} \sim 0.9 \times 10^{-38}\,{\rm cm}^2$~\cite{Monroe}.
Thus we expect to produce 
$\sim 2\times (2\times 10^5)\sigma(700\,{\rm MeV})/\sigma_{QE} 
\sim 140\;(g_\omega/10)^4$ events.  
The extra factor of $2$ arises from the fact that our process 
involves both $n$ and $p$ while CCQE involves only $n$ in a carbon target. 
Our cross-section has a distinctly flat angular distribution,
$d\sigma/d\cos\theta_\gamma = \sigma(E_\nu )/2$
where $\cos\theta_\gamma = \vec{p}\cdot\vec{q}/E_\nu E_\gamma$. 
We note, however, that the photon will be pulled forward
as form-factor effects, which we have ignored, set in.

While this estimate is extremely naive, it is
encouraging that existing experiments 
may already have sensitivity
to our effect.  For a more refined analysis 
we require an improved $\sigma(E_\nu)$ on
nuclear (carbon) targets, including
coherence and form-factor effects, and convolution
of $\sigma(E_\nu)$ with the beam spectrum.  

Additional anomaly mediated interactions 
can also be studied systematically starting from the WZW term.
For instance, a term 
$\sim \epsilon_{\mu\nu\rho\sigma} \partial^\mu \pi^0 Z^\nu F^{\rho\sigma}$ 
involves pion exchange with the nucleus in place of $\omega$ exchange.  
However, this process contains 
$1/f_\pi^4 < g_\omega^4/m_\omega^4$, and is not the dominant effect. 
Also, in contrast to (\ref{one}),  
the amplitude for this process does not add coherently 
over individual nucleons at low energy~\footnote{
Coherent enhancements of neutrino-photon interactions
were studied in a current algebra framework in: 
D.~Rein and L.~M.~Sehgal,
Phys.\ Lett.\  B {\bf 104}, 394 (1981)
[Erratum-ibid.\  B {\bf 106}, 513 (1981)].
However, at the energies of interest, 
the assumption of coherence on the entire nucleus 
leads to only small effects confined to the forward direction.
}.

We thus conjecture that the anomaly mediated 
process may be relevant to the MiniBooNE excess.
Moreover, it could lead to multiple enhanced prospects for observing
quasi-elastic neutrino scattering on nuclei. 
Modern neutrino experiments could ultimately provide a
normalization of the uncertain input parameter,
$g_\omega$.
While the anomaly mediated neutrino process is higher order, it contains
a hard photon. Even for relatively low energy reactor
neutrinos, $E_\nu \sim 3\,{\rm MeV}$, the enormous flux availability
suggests that the anomaly process may be observable. The hard photon
also provides a possible observable for a quartz-Cerenkov detector, 
or liquid halogen bubble detectors, \etc.

\section{Neutron Star Cooling}

The anomaly mediated process is competitive with the 
conventional processes for neutron star
cooling. For illustration, we consider a particular coherent
subprocess in a superconducting neutron star core, but the 
anomaly mediated process will
have a more general applicability.
Neutron star cooling is reviewed in  \cite{Shapiro,Raffelt,Yak,Pethick}. 

The full set
of processes contributing to neutron star
cooling is complex.
In the ultra-high
density inner core the energy loss is dominated by the direct Urca process,
\begin{equation}
n \rightarrow p + e^- + \bar \nu_e, \qquad e^- + p \rightarrow n + \nu_e \,. 
\end{equation}
Throughout most of star the nucleons are at lower densities,
$\lesssim 10^{15}\,{\rm g}\, {\rm cm}^{-3}$, and are
Fermi degenerate.  In this case, the direct Urca process is highly suppressed.
Energy is then typically lost 
by the modified Urca (mUrca) process where 
a bystander nucleon is included to conserve
energy and momentum. 
The mUrca process is affected by the superfluid phase, which
appears below a critical temperature $T_c \sim 10^{10}\,{\rm K}$. 
This reduces the mUrca rate.  However, new cooling mechanisms 
associated with Cooper pairing of nucleons may turn on at
$T\lesssim 10^{9}\,{\rm K}$, compensating the mUrca suppression.  

We presently consider typical neutron star
densities of $2\rho_0 = 5.6\times 10^{14}\,{\rm g}\,{\rm cm}^{-3}$
(twice nuclear density $\rho_0$)
and assume that the regime $10^9 \lesssim T \lesssim 10^{10}\,{\rm K}$
contains the standard mUrca processes. 
We summarize these effects by (see \eg, Table 2 of \cite{Yak}):
\begin{equation}
\label{thirteen}
Q_\nu^{\rm mUrca} = (10^{18}\; - \;10^{21}) \times (T_9)^{8}\; {\rm erg\;s^{-1}\; cm^{-3}} \,,
\end{equation}
at temperature $T = T_9 \times 10^9 K$. This describes
a collection of various processes, ignoring superfluidic suppression, and 
permits a naive comparison to our process.

We now want to estimate the cooling rate due to the anomaly
mediated process.
Our basic assumptions are:
\begin{enumerate}
\item The degenerate protons pair to make a superconductor. 
This spontaneously breaks $U(1)_{EM}$ and gives a mass to the photon, 
also known as the inverse penetration length of the superconductor.
The mass depends sensitively on strong interaction models, 
we take $m_{\gamma} \sim 1 \, {\rm MeV}$
as a typical value.

\item The (massive) photons are in thermal equilibrium 
with the neutrons, protons and electrons
in the superconducting interior of the neutron star
and we work in the limit $T \lesssim m_\gamma$.
\end{enumerate}

The photons are therefore 
characterized by the phase-space
distribution function and number density,
\begin{equation}
f(p_\gamma) = \left[ e^{E_\gamma /T}-1 \right]^{-1} \,,
\quad 
n_\gamma = g \int \frac{d^3p_\gamma }{(2 \pi)^3} f(p_\gamma) \,,
\end{equation}
with $g=3$ for a massive spin-one particle. 
For $T\lesssim 10^{10}\,{\rm K} \sim m_\gamma$,
we can approximate $E_\gamma \sim m_\gamma + |\vec{p}_\gamma|^2/2 m_\gamma$. 
The emissivity of neutrinos
due to $\gamma \rightarrow \nu \bar \nu$ is
\begin{equation}
Q_\nu^{\rm anom} = 
3 \int \frac{d^3p_\gamma}{(2 \pi)^3} ~ E_\gamma ~ 
\Gamma_{p_\gamma}(\gamma \rightarrow \nu \bar \nu)~ f(p_\gamma),
\end{equation}
thus requiring the decay rate of the photon,
$\Gamma_{p_\gamma}$. 

In the star rest frame the baryon number 
current is $J_\mu =(n_B, \vec{0})$, 
and the photon 4-momentum is 
$p_\gamma \approx (m_\gamma, \vec{p}_\gamma)$. 
However,
it is convenient to work
in the photon rest-frame, where
$p_\gamma^\prime =(m_\gamma, \vec{0})$. 
In this frame the nucleon current is
slightly boosted: $J_\mu^\prime = n_B \eta_\mu$, where 
$\eta_\mu \approx (1, \vec{\beta})$ and   
$\vec{\beta} \approx -\vec{p}_\gamma/m_\gamma$. 
In the nonrelativistic limit we thus have:
\bea
\Gamma_{p_\gamma} &=& \frac{\kappa_a^2n_B^2}{2m_\gamma}
\int\frac{d^3p_1}{2E_{1}(2\pi)^3}
\frac{d^3p_2}{2E_{2}(2\pi)^3} \nonumber \times \\
&& \qquad |M|^2 (2\pi)^4\delta^4(p_\gamma^\prime - p_1-p_2) \,, 
\eea
and the invariant final-state spin-summed and 
initial state spin-averaged matrix element is:
\beq
|M|^2 = \frac{2 m_\gamma^2 }{3}
\epsilon^{\mu\nu 0}_{\quad\;\, \rho}\, \epsilon^{\alpha\beta 0\rho }
\eta_\mu\eta_\alpha
\Tr[(1-\gamma^5)\gamma_\nu  \slash{p}_1 \gamma_\beta\slash{p_2}] \,. 
\eeq
This results in the decay rate:
\bea
\Gamma_{p_\gamma} & = &  
\frac{2m_\gamma \kappa_a^2n_B^2}{9\pi}
|\vec{p}_\gamma|^2  \,, 
 \eea
and emissivity:
\bea
\label{lum}
Q_\nu^{\rm anom} &=& Ce^{-m_\gamma/T}(m_\gamma T)^{5/2}
\frac{G^2_F m^2_\gamma n_B^2 }{m_\omega^4} \,, 
\nonumber \\
\quad C &=& 
\frac{\sqrt{2\pi}\alpha g_\omega^4 }{16\pi^{6} }= 0.012 \; - \; 0.96 \,. 
\eea
The range of $C$ corresponds to a range of $g_\omega = 10.0$ to
$g_\omega = 30.0$. We note that
both $g_\omega$ and $m_\omega$ run
with nuclear density, \eg, Refs.~\cite{Caillon}
obtain $m_\omega$ reduced by a factor $\sim 0.6 \; - \;0.8$ 
at $\rho \gtrsim \rho_0$. Holding $m_\omega$ fixed,
larger values of $g_\omega$ account for this effect, 
as well as other effects, \eg, 
higher resonance contributions.
For standard density $2\rho_0 =5.6\times 10^{14}\,{\rm g}\,{\rm cm}^{-3}$, 
this yields 
an emissivity of
\bea
Q_\nu^{\rm anom} &=& 2.31 \times 10^{22} 
\, {\rm erg}\; {\rm s}^{-1}\; {\rm cm}^{-3} 
\times \nonumber \\
&& \quad 
m^{9/2}(g_\omega/10)^4
e^{(-11.6m/T_9)}(T_9)^{5/2} \,,
\eea
where $m= m_\gamma/(1\,{\rm MeV})$.

\begin{figure}[t]  
\vspace{2.5cm}  
\includegraphics{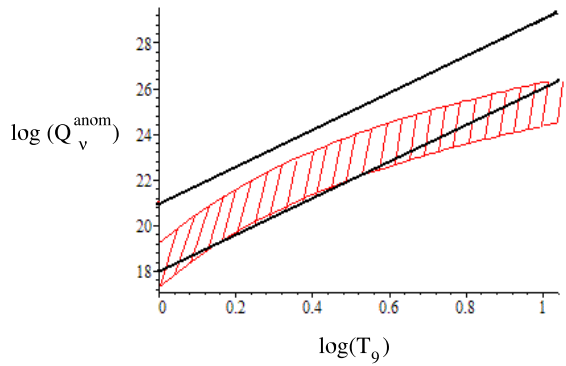}  
\vspace{2.0cm}  
\label{cool}  
\caption{ 
$\log(Q_\nu^{\rm anom})$, with $Q$ measured in 
${\rm erg}\,{\rm s}^{-1}\, {\rm cm}^{-3}$, versus
$\log(T_9)$ for the range $g_\omega = 10\; - \; 30$ (hatched)
compared to the range of standard mUrca processes 
of eq.(\ref{thirteen}).  
The curves for mUrca do not include superfluidic suppression factors.
}  
\vspace{-0.25cm}
\end{figure}  

We see in Fig.(2) that this is competitive with 
the cooling rate from mUrca processes~\footnote{
Although our effect is one-loop order, 
it is significantly enhanced 
relative to other loop processes, \eg,
to the electromagnetic penguin, $\gamma\rightarrow \bar{\nu}\nu$, 
by a factor of 
$\sim T g_\omega^4 n^2_B/m_\omega^4 m_\gamma^3(\ln(M_W/m_\gamma))^2 \sim 10^3 $.}.
We note that in the early phase of neutron star formation 
with $T \gtrsim 10^{11} \,{\rm K}$, our process may actually dominate.
This will be developed and reported elsewhere~\cite{Harvhill2}.

\section{Conclusions}

Anomaly mediated 
interactions between photons and neutrinos at finite baryon
density may play an important role in laboratory
neutrino experiments and astrophysical processes.
The new interaction (\ref{one}) may be relevant in accounting for the
MiniBooNE low energy excess. It may also play a significant
role in neutron star cooling and early stage evolution.  There are
many potentially important
applications in various other physical regimes.
We will present a
more detailed analysis and discussion elsewhere,
including the detailed derivation of pCS and axion interactions
from the WZW term~\cite{Harvhill2}\cite{Son:2004tq}.

We further remark that 
the axion will have a similar induced 
coupling to the photon and the $\omega$, leading to an interaction
of the form:
\beq
 c_{axion}\frac{ eN_c}{24\pi^2}\frac{g_\omega^2}{m_\omega^2}\epsilon_{\mu\nu\rho\sigma} \frac{\partial^\mu a}{f_a}
F^{\nu\rho}\bar{N} \gamma^\sigma N \,, 
\eeq
where $c_{\rm axion}$ is calculable from a given axion model. 
An important application is to consider axion emission and 
the resulting bounds on axion couplings from supernovae (SN1987A).

\vskip 0.2in
\noindent
{\bf Acknowledgements}
\vskip 0.1in
\noindent
We thank W. Bardeen, J. Conrad, and P. Cooper 
for helpful discussions.
Research supported by the U.S.~Department of Energy  
grant DE-AC02-76CHO3000 and by NSF Grants PHY-00506630 and 0529954.
This work was supported in part by the Joint Theory Institute funded by 
Argonne National Laboratory and the University of Chicago.
JH acknowledges the hospitality of the Aspen Center for Physics during part
of this work.


\end{document}